\documentclass[a4paper,fleqn,usenatbib,onecolumn]{mnras}

\usepackage[T1]{fontenc}
\usepackage{ae,aecompl}
\pdfoutput=1 

\usepackage{graphicx}   
\usepackage{amsmath}    
\usepackage{amssymb}    

\def\ep{\varepsilon}
\newcommand\equ[1]{{\rm (\ref{#1})}}
\def\integer{{\mathbb Z}}
\def\positive{{\mathbb N}}
\newcommand\beqa[1]{ \begin{eqnarray} \label{#1}}
\newcommand{\eeqa}{ \end{eqnarray} }
\newcommand\beq{ \begin{equation}}
\newcommand{\eeq}{ \end{equation} }

\newcommand{\beqano}{ \begin{eqnarray*} }
\newcommand{\eeqano}{ \end{eqnarray*} }
\usepackage[usenames]{xcolor}

\title[Secondary resonances in the spin-orbit problem]{The theory of secondary resonances in the spin-orbit problem}

\author[I. Gkolias et al.]{
Ioannis Gkolias,$^{1}$\thanks{E-mail: gkolias@mat.uniroma2.it}
Alessandra Celletti,$^{1}$
Christos Efthymiopoulos,$^{3}$
Giuseppe Pucacco$^{2}$
\\
$^{1}$Department of Mathematics, University of Rome Tor Vergata,Via della Ricerca Scientifica 1,
00133 Rome, Italy \\
$^{2}$Department of Physics, University of Rome Tor Vergata,Via della Ricerca Scientifica 1,
00133 Rome, Italy\\
$^{3}$Academy of Athens,Research Center of Astronomy and Applied Mathematics,Soranou Efessiou 4, GR-11527 Athens, Greece
}

\date{Accepted XXX. Received YYY; in original form ZZZ}

\pubyear{2016}

\begin{document}
\label{firstpage}
\pagerange{\pageref{firstpage}--\pageref{lastpage}}
\maketitle

\begin{abstract}
We study the resonant dynamics in a simple one degree of freedom, time dependent Hamiltonian model describing spin-orbit interactions. The equations of motion admit periodic solutions associated with resonant motions, the most important being the synchronous one in which most evolved satellites of the Solar system, including the Moon, are observed. Such \sl primary \rm resonances can be surrounded by a chain of smaller islands which one refers to as \sl secondary \rm resonances. Here, we propose a novel canonical normalization procedure allowing to obtain a higher order normal form, by which we obtain analytical results on the stability of the primary resonances as well as on the bifurcation thresholds of the secondary resonances. The procedure makes use of the expansion in a parameter, called the detuning, measuring the shift from the exact secondary resonance. Also, we implement the so-called `book-keeping' method, i.e., the introduction of a suitable separation of the terms in orders of smallness in the normal form construction, which deals simultaneously with all the small parameters of the problem. Our analytical computation of the bifurcation curves is in excellent agreement with the results obtained by a numerical integration of the equations of motion, thus providing relevant information on the parameter regions where satellites can be found in a stable configuration.
\end{abstract}

\begin{keywords}
Spin-orbit problem -- resonances -- normal form
\end{keywords}



\section{Introduction}
The investigation of the rotational dynamics of natural bodies, either satellites or planets, has
remarkable consequences, which go well beyond the description of the rotational state of the body.
For example, from the analysis of the rotational dynamics, one can infer the internal composition
of the body (liquid/solid core, liquid/solid mantle, etc. \cite{Mar2007,Taj2014,Tho2016}). The description of the rotational motion must take into account
possible resonant coupling states with the orbital motion. As it is well known, most of the
evolved satellites of the Solar system are trapped in a synchronous (or 1:1) spin-orbit resonance,
which implies an equality between the rotational period of the satellite and its orbital period
about the host planet.

A model describing the synchronous resonance of a satellite under the gravitational influence of a central planet
is the so-called \sl spin-orbit problem \rm (\cite{goldpeale,zamp1}). In the model's most basic form, we assume that a triaxial satellite orbits on a Keplerian ellipse, that its spin-axis is perpendicular to the orbital plane and that it coincides with the smallest axis of the satellite. The model is ruled by two parameters: the asphericity of the
satellite and the orbital eccentricity.

The equations of motion of the spin-orbit problem admit periodic solutions, which correspond to
resonant motions associated with a commensurability between the
rotational and orbital periods. The synchronous resonance is a special case of such commensurability which we refer to as a \sl primary \rm 1:1 spin-orbit resonance.
Beside the synchronous resonance, a special role is played by the 3:2 resonance, which is the
only non-synchronous resonance observed in the Solar system: twice the orbital period of Mercury
is equal to three times its rotational period.

The spin-orbit primary resonances are surrounded by librational regions with size depending on
the values of the parameters (asphericity and eccentricity). The librational region may contain
chains of \sl secondary \rm islands of different period. An interesting study of secondary resonances
of the spin-orbit problem is presented in \cite{Wisdom}, which provides an analytical model for
the 3:1 secondary resonance of the 1:1 primary resonance. A low order estimate of the bifurcation
threshold of the 3:1 secondary resonance is computed and an application to Enceladus is presented,
showing that the trapping in the 3:1 secondary resonance might cause tidal heating (see also \cite{BM}
for a study of new possibilities of stable spin-orbit configurations which appear in binary systems when both bodies are aspherical).

The aim of this work is to present
an analytical technique to study the secondary resonances of the spin-orbit problem. Specifically,
we compute an integrable approximation of the original model by constructing a suitable higher order normal
form\footnote{We remark that a second order normal form construction around
the 1:1 primary resonance was computed in \cite{FS}, although such analytical construction is not
adequate to reconstruct the dynamics around secondary resonances.}.
This construction allows us to study stability of the primary resonances
and bifurcation properties of families of periodic orbits associated with secondary resonances.

Two basic tools are used to obtain the results:

i)We explore the idea of {\it detuning} the frequency ratio of the secondary resonance with the orbital frequency. This means to introduce a small parameter $\delta$ measuring the difference between the value of the above frequency ratio for a body with specific physical parameters (see section 3) and the exact resonant value. In our approach $\delta$ is carried along all series expansions as one more small parameter (the other natural two ones are the eccentricity of the orbit and the distance from the exact primary resonant state). The effectiveness of the detuning technique has been demonstrated in other nearly-resonant dynamical systems (see \cite{MP13,MP14}). As shown below, it provides a powerful way to analytically study the spin-orbit secondary resonances in a range of parameter values quite relevant to true astronomical objects.

ii) The detuning method is combined with a method called `book-keeping' (see \cite{CHR}). Essentially, this means to use a unique symbol in order to represent all the small parameters in the series according to a `book-keeping rule', i.e., a symbolic rule expressing their relative size in the problem. The relevance of the book-keeping method manifests itself in high-order normal form computations performed, as below, with the aid of a computer program. In fact, our whole approach is designed so as to be easily transferable to a computer program.

As a result, our normal form construction manages to capture, in a very accurate way, all the relevant features of the original system. In particular, with our construction we are able to compute analytically and provide:

i) The position and time evolution of the primary resonant state. As a particular application, we give analytical formulas for the most important of these states, namely the synchronous one. This is represented by an equilibrium point in the normal form, which, when back-transformed to the original state variables of the rotating body, yields a periodic orbit. An analytic expression for this orbit is provided in section 3 below.

ii) An analytical representation of the phase space portrait around the primary resonance. This is obtained from back-transforming to the original variables the phase space portrait of the normal form variables; the latter is obtained trivially, since it represents a system of one degree of freedom. This procedure allows to analytically predict the changes of stability character of the primary resonance associated with bifurcations of the secondary resonances, the position and width of secondary resonances, and the related change of topology in the phase space.

iii) The analytic determination of the bifurcation limit for secondary resonances, exemplified below by the case of the 2:1 secondary resonance. A detailed numerical computation of the bifurcation curves was already presented in \cite{Melnikov} and \cite{MS} (we refer also to \cite{Bruno} for a thorough study of families of periodic orbits in the spin-orbit problem). Here, instead, we give an analytical theory for those results, allowing to draw conclusions about the stability regions (in the parameter plane of the asphericity and eccentricity) of the primary
resonance as well as about the mechanism of transition to instability (compare with \cite{MS}).

We validate our methods in two ways: 1) by directly comparing our results with high precision numerical computations and 2) by giving error estimates of the accuracy of our normal form construction, computing the size of the `remainder' of the normal form (see section \ref{sec:modelsecondary}).

As shown below, the parameter domain within which our results apply covers a large set of physical systems. Most minor planetary satellites and moonlets have irregular shapes and some have been conjectured to fall in the unstable synchronous domain beyond the bifurcation threshold, a fact that might imply a chaotic rotation for such bodies (\cite{Wis1984,Wis1987,MS}). However, as discussed in section \ref{sec:discappl}, recent observations (\cite{Pravec}) suggest the opposite to be true for synchronous binary asteroid systems. Providing analytical formulas allows to address this issue and to constrain the domain in parameter space where natural bodies can be observed in a stable configuration.  This could also be helpful in other contexts, like in the study of the dynamics of exoplanetary systems.

This paper is organized as follows. The model for the study of spin-orbit resonances is introduced in Section~\ref{sec:model};
an analytical theory for the 2:1 secondary resonance of the synchronous state is given in Section~\ref{sec:modelsecondary} followed by a series of useful computations and the validation of our method. As another example, in Section~\ref{sec:secondary32} our method is applied to the 2:1 secondary resonance of the 3:2 primary one. Finally, in Section \ref{sec:discappl} we summarize our results and discuss their application in natural systems.

\section{Model}\label{sec:model}
We consider a triaxial satellite moving around a central body on a fixed Keplerian ellipse. We assume that the spin-axis of the satellite is perpendicular to the orbit plane and that it coincides with the smallest physical axis. Dissipative forces, e.g. due to tidal effects, are not considered. We note that dissipative effects are important in driving the system to particular resonant configurations (see \cite{Fer2008,Fer2015} and references therein). Here, however, we are interested in identifying which configurations are stable under conservative dynamics, in order to characterize the possible endstates of spin-orbit evolution.

Under the above assumptions, the model is described by the following differential equation (\cite{goldpeale,zamp1,MD,Alebook}):
\begin{equation}
\ddot{\theta}+n \frac{\ep^2}{2} \left(\frac{a}{r(t)}\right)^3 \sin(2\theta-2f(t))=0\ ,
\label{eq:soeom}
\end{equation}
where $\theta$ is the angle formed by the largest physical axis of the satellite and the periapsis line, $n$ is the orbital frequency, $a$ is the semi-major axis, $f=f(t)$ denotes the true anomaly and $r=r(t)$ is the distance between the satellite and the central body. The parameter $\ep$ is related to the shape of the satellite and it is calculated from the relation:
$$
\ep = \sqrt{\frac{3 (B-A)}{C}}
$$
with $A<B<C$ the moments of inertia of the satellite. This parameter, also called asphericity, gives us the degree of divergence from a perfectly spherical shape.

The units in \eqref{eq:soeom} can be chosen such that $n=a=1$, which implies that the orbital period is exactly equal to $2 \pi$.
The time evolution of the true anomaly can be calculated from the second Kepler law:
\begin{equation}
\dot{f}=\frac{1}{p^{3/2}}(1+e \cos f)^2\ ,
\label{ft}
\end{equation}
where $p = (1 - e^2)$ is the semi-latus rectum of the Keplerian ellipse and $e$ is its eccentricity. The time evolution of the distance from the perturber can the be computed as a function of the true anomaly:
\begin{equation}
r(t)=\frac{p}{1+e \cos(f(t))}\ ,
\label{rt}
\end{equation}
where $f(t)$ is obtained by integrating \equ{ft}.

Equation \eqref{eq:soeom} is associated with a one--dimensional, time--dependent Hamiltonian function of the form:
\begin{equation}
H(p_{\theta},\theta,t)=\frac{p_{\theta}^2}{2}-\frac{\ep^2}{4} \frac{1}{r^3(t)}\cos(2 \theta-2f(t))\ .
\label{eq:fullham}
\end{equation}
We remark that \equ{eq:fullham} is integrable in two cases:
\begin{itemize}
    \item[$(i)$] when $\ep=0$, namely when the satellite is axisymmetric with $A=B$. In this case, the torque exerted on
    the satellite vanishes, leading to a free rotation. The Hamiltonian
    \equ{eq:fullham} reduces to $H(p_{\theta},\theta,t)=\frac{p_{\theta}^2}{2}$, so that $p_{\theta}$ is constant
    and $\theta$ becomes a linear function of the time;
    \item[$(ii)$] when the orbit is circular, namely the eccentricity is equal to zero, which leads to have $r$ constant
    from \equ{rt} and $f$ coinciding with time due to \equ{ft}. Hence, one can introduce a rotating frame with frequency equal
    to the orbital frequency, so that the corresponding dynamics reduces to that of a pendulum. Note that in this case the synchronous rotation is the only endstate for the roto-translational configuration in the presence of typical
dissipative forces, for example as those associated with a tidal torque.
   \end{itemize}

When $\ep$ or $e$ are not zero, the Hamiltonian system is non-integrable, providing a full suite of nonlinear phenomena.
Among these phenomena, we are interested in the bifurcations of the periodic solutions which correspond to the so-called \sl spin-orbit resonances, \rm which occur whenever the periods of revolution around
the planet and of rotation about the spin-axis are rationally dependent, namely there exists $m,k\in\integer$, $k\not=0$, such that
$$
{{\dot\theta}\over{n}}={m\over k}\ .
$$
We shall refer to these resonances as \it m:k \sl primary \rm spin-orbit resonances. In particular, we will focus on the 1:1 and 3:2 resonances. The first one, also called \sl synchronous \rm resonance, is very common in the Solar system and it implies that the periods of rotation and revolution of the satellite are the same. The Moon and most of the large natural satellites are trapped into a nearly synchronous rotation. The only body in our solar system observed in a 3:2 spin-orbit resonance is  Mercury: twice its period of revolution equals three times its period of rotation.

In the sequel, we are interested in studying the bifurcation phenomena associated with the appearance of a chain of \sl secondary \rm islands within the librational region around the center of the primary resonance. A mathematical description of primary and secondary resonances will be given in Section~\ref{sec:model}. Bifurcation curves in the plane of the two control parameters ($e,\ep$) will be computed and compared with those produced by numerical computations of stability thresholds based on the trace of the monodromy matrix.

\section{Analytical modelling of secondary resonances}\label{sec:modelsecondary}

In this Section we introduce an analytical method to investigate secondary resonances around a given primary one.
This technique is based on the construction of a suitable resonant normal form. Since we aim at reaching good accuracy in the analytical predictions, we will pay special attention to develop a procedure of normalization to arbitrary order.
The computations will be detailed for the 2:1 secondary resonance of the 1:1 primary. With the help of the resonant normal form, we perform a series of computations. First we investigate the phase space through Poincar\'e surfaces of section, showing that our construction not only captures the basic dynamics of the system but also represents the motion around the primary resonance in a very accurate manner. Then, we compute the characteristic curve of the synchronous resonance and compare it with numerical results. Finally, we conclude this section with a discussion on the estimation for the error of our method.

\subsection{Primary Resonance: The synchronous case}\label{sec:primary}
The analytical treatment of the problem starts by making explicit the dependence of (\ref{eq:fullham}) on time.
Due to the assumption that the satellite moves on a Keplerian orbit, both the true anomaly and the orbital radius
are known functions of the time through \equ{ft} and \equ{rt}, respectively. The solutions
$f=f(t)$ and $r=r(t)$ can be expanded in Fourier series. As a consequence, the spin-orbit Hamiltonian takes the form:
\begin{equation}
H(p_{\theta},\theta,t)=\frac{p_{\theta}^2}{2}-\frac{\ep^2}{4} \sum_{m \neq 0,m=-\infty}^{m=\infty} W\left(\frac{m}{2},e\right) \cos(2 \theta- m t )\ ,
\label{eq:fullhamfour}
\end{equation}
where the coefficients $W=W\left(\frac{m}{2},e\right)$ are the classical $G$ functions of \cite{Kau1966} and they are series in the eccentricity of order $e^{|m-2|}$ (\cite{Cay1861}, see also \cite{MD,Alebook}):
\beqano
W\left(\frac12,e\right)&=&-\frac{e}{2}+\frac{e^3}{16}+\ldots,\nonumber\\
W(1,e)&=&1-\frac{5}{2}e^2+\frac{13}{16}e^4+\ldots,\nonumber\\
W\left(\frac32,e\right)&=&\frac{7}{2}e-\frac{123}{16}e^3+\ldots
\eeqano

We then find it convenient to introduce the extended phase-space to get rid of the time-dependence
of the Hamiltonian \equ{eq:fullhamfour}. To this end, we introduce a \sl dummy \rm action $p_2$,
conjugated to the time variable with frequency equal to the orbital frequency. By means of the extended phase-space transformation
$$
(p_{\theta},H,\theta,t) \rightarrow (p_1,-p_2, q_1,q_2)\ ,
$$
we obtain the transformed null Hamiltonian ($H_E=0$) given by
$$
H (p_1,p_2,q_1,q_2)=\frac{p_1^2}{2}+p_2-\frac{\ep^2}{4} \sum_{m \neq 0,m=-\infty}^{m=\infty} W\left(\frac{m}{2},e\right)
\cos(2 q_1 - m q_2 )\ .
$$

Our goal now is to study the small-amplitude oscillations around the primary resonances of the system. In order to do so,
we apply a canonical transformation to a rotating frame introducing a so-called \sl resonant angle. \rm
More precisely, we consider the change of coordinates
$$
 (p_1,p_2, q_1,q_2) \rightarrow (p_{\psi},p_{\phi},\psi,\phi)
$$
with
\beqa{rotransf}
 p_1 &=& p_{\psi} + \frac{m}{k}\ ,\nonumber\\
 p_2 &=& p_{\phi} - \frac{m}{k} p_{\psi}\ ,\nonumber\\
 \psi &=& q_1 - \frac{m}{k} q_2\ ,\nonumber\\
 \phi &=& q_2 \label{rotransf1}
\eeqa
for some $m$, $k$ integers.
The Hamiltonian in the transformed variables turns out to be
\begin{equation}
H=p_{\phi}+\frac{p^2_{\psi}}{2}-\frac{\ep^2}{4} W_0(m,e) \cos(2 \psi) + H_{\text{pert}}(\psi,\phi;e,\ep)\ ,
\label{HR}
\end{equation}
where  $W_0(m,e)$ is the term with the lowest order in the eccentricity for the considered resonance and
$H_{\text{pert}}$ represents all other coefficients, transformed under the change of coordinates \equ{rotransf}. The ratio $m/k$ in \equ{rotransf1} is chosen according to which primary resonance we are interested to explore. We remark that the Hamiltonian \equ{HR} is split into an integrable part and a perturbing part: the integrable part is the sum of the dummy action $p_{\phi}$ and a pendulum-like Hamiltonian in the resonant angle $\psi$.

For the synchronous (1:1) resonance the transformation (\ref{rotransf}) reads as
$$
p_1=p_\psi +1\ , \quad p_2=p_{\phi}-p_{\psi}\ , \quad \psi = q_1 - q_2\ , \quad \phi = q_2\ .
$$
Since we are interested in small amplitude oscillations around the synchronous state, we will expand the angle $\psi$
in Taylor series as $\cos(2 \psi) = 1-2 \psi^2 + \ldots$ up to an order that is consistent with the rest of our normal form construction, so to get the expanded Hamiltonian:
\begin{equation}\label{eq:hamprim11}
H=p_{\phi}+\frac{p^2_{\psi}}{2}+\frac{\ep^2}{2} \psi^2 + H_{\text{pert}}(\psi,\phi;e,\ep)\ .
\end{equation}

\subsection{Secondary Resonance: The 2:1 Case. Detuning}\label{sec:secondary11}

It is important to notice that, in the Hamiltonian (\ref{eq:hamprim11}), the frequency of the small amplitude oscillations around the primary resonance is equal to $\ep$. A \sl secondary resonance \rm will appear when this frequency becomes commensurable with the orbital frequency which, in the current units, is identically equal to 1. Therefore, for an $\ell$:{\it k} secondary resonance to occur, the value of $\ep$ should be close to $k/\ell$. To quantify the deviation from the exact resonance, we introduce a {\it detuning} parameter $\delta$ (see \cite{Vf}), such that:
$$
\delta \equiv \ep - \frac{k}{\ell}.
$$

For example, let us consider the case of the 2:1 secondary resonance, for which we introduce the detuning parameter as
$$
\delta=\ep-\frac12\ ,
$$
while the associated expanded Hamiltonian is given by
\begin{equation}\label{HH21}
H=p_{\phi}+\frac{p^2_{\psi}}{2}+\frac{1}{8} \psi^2 +  H_{\text{pert}}(\psi,\phi;e,\delta)\ .
\end{equation}
The integrable part of the Hamiltonian corresponds now to a pair of a rotator and a  harmonic oscillator with \sl unperturbed \rm frequencies $\omega_1=1, \omega_2=1/2$ respectively, whereas the \sl small \rm term of the form $(1/2) \delta \psi^2$ is now treated as a higher-order term and put inside the perturbation. Next, we introduce the action-angle variables for the integrable part, say
\begin{equation}\label{AARV}
\psi = \sqrt{\frac{2 J}{\omega_2}} \sin{u}\ , \quad p_{\psi} = \sqrt{2 J\omega_2} \cos{u}\ , \quad J_{\phi} = p_{\phi}\ ,
\end{equation}
which brings our Hamiltonian into the following form:
$$
H=J_{\phi}+ \frac{1}{2} J + H_{\text{pert}}(J,u,\phi;e,\delta)\ .
$$
The perturbing part $H_{\text{pert}}$ is a Fourier series in $u,\phi$ of the form
\begin{equation}\label{FEX}
H(J,u,\phi;e,\delta) = \sum_{k_0,k_1,k_2} c_{k_0k_1k_2} (e,\delta) J^{k_0\over 2} {\rm e}^{i (k_1 u + k_2 \phi)}\ , \quad k_0,k_1,k_2 \in \mathbb N\ ,
\end{equation}
which in general does not have the D'Alembert character (\cite{mho}).

\subsection{Book-keeping}\label{sec:book}
The Hamiltonian (\ref{FEX}) contains three different types of small quantities whose powers will appear in subsequent series expansions: besides the parameters $e$ and $\delta$, there is also the action variable $J$. According to the definitions of Eq. (\ref{AARV}), the orbits with $J\neq 0$ are orbits in the neighborhood of the synchronous state which librate with an amplitude ${\cal O}(J^{1/2})$. In order to deal simultaneously with all three different small quantities appearing in the problem, a convenient way is to introduce a \sl book-keeping \rm parameter $\lambda$ (see \cite{CHR}); this scaling parameter determines the ordering of different terms in \eqref{FEX}: the size of each term decreases in as much the powers of $\lambda$ increases. More specifically, in the Hamiltonian (\ref{FEX}) we make a set of substitutions, called `book-keeping rules', selected as follows:
\begin{eqnarray}
 e &\rightarrow &\lambda e \label{booke}\\
 \delta &\rightarrow& \lambda \delta\ \label{bookd}\\
 J^q &\rightarrow& \lambda^{2q-2} J^q\ .  \label{bookJ}
\end{eqnarray}
In the above expressions, $\lambda$ is a symbol carried along in all the expansions, whose numerical value is set to 1 in the end of the normalization process. Thus, quantitatively, $\lambda$ plays no role. However, all the expansions, series, transformations, etc., are from now on considered in powers of $\lambda$. This enables a very general ordering scheme, so that we are in the position to apply a canonical normalization approach in which both internal and control parameters determine the hierarchy in the normal form. Let us note that the rules \eqref{booke} and \eqref{bookd} attach the same order of smallness to eccentricity and detuning: other choices are possible, but these appear to comply the most with the size of the small parameters in our domain of interest. On the other hand, the choice of the book-keeping rule \eqref{bookJ} reflects a natural scaling of the libration phase-space variables going with half-integer powers of the action variable $J$.

After introducing $\lambda$, the `book-kept' Hamiltonian reads as
\begin{equation}\label{eq:hambookkept}
H=J_{\phi}+ \frac{1}{2} J + \lambda H_1(J,u,\phi;e,\delta) + \lambda^2 H_2(J,u,\phi;e,\delta) + \ldots
\end{equation}

\subsection{Canonical normalization}\label{sec:norm}
The Hamiltonian (\ref{eq:hambookkept}) contains already the information about the phase space structure, bifurcation of secondary resonances, stability, etc. This is embedded in Hamilton's equations of motion for the variables $(J,J_\phi,u,\phi)$. In order to unravel this information, the core of the normal form analytical approach is to find a transformation from the set $(J,J_\phi,u,\phi)$ to new variables in which the dynamics becomes transparent. To this end, we now apply the Hori-Deprit approach based on Lie series (\cite{gior}).

In this method, we seek to find a sequence of consecutive near-identical canonical transformations
\begin{equation}\label{eq:transvarlie}
(J,J_\phi,u,\phi) \equiv (J^{(0)},J_\phi^{(0)},u^{(0)},\phi^{(0)}) \rightarrow (J^{(1)},J_\phi^{(1)},u^{(1)},\phi^{(1)}) \rightarrow (J^{(2)},J_\phi^{(2)},u^{(2)},\phi^{(2)}) \ldots
\end{equation}
such that, after $n$ steps, the Hamiltonian, transformed in the new variables, takes the form
\beq\label{hamnf}
H^{(n)}=Z_0+\lambda Z_{1}+\ldots+\lambda^{n} Z_{n} + \lambda^{n+1} H^{(n)}_{n+1} + \lambda^{n+2} H^{(n)}_{n+2}+\ldots
\eeq
The quantity
$$
Z^{(n)}=Z_0+\lambda Z_{1}+\ldots+\lambda^{n} Z_{n}
$$
is called the normal form. The sequence of transformations (\ref{eq:transvarlie}) is designed so as to yield a normal form function $Z^{(n)}$ whose dynamics is easy to analyze. Then, back-transforming to the original variables, we obtain an approximation of the dynamics of the original system as well. On the other hand, the quantity
\beq\label{rem}
R^{(n)}=\lambda^{n+1} H^{(n)}_{n+1} + \lambda^{n+2} H^{(n)}_{n+2}+\ldots\ ,
\eeq
called the remainder, serves to obtain estimates of the error of the $n$-th step normal form approach, i.e., the difference between the true dynamics and the dynamics under the normal form term $Z^{(n)}$ only. In the Hori-Deprit approach, the sequence of transformations (\ref{eq:transvarlie}) is obtained by a sequence of Lie generating functions $\chi_1$, $\chi_2$, ... Namely, the transformations are given by:
\begin{eqnarray}\label{lietra}
~J&=&\exp(L_{\chi_{n}})\exp(L_{\chi_{n-1}})\ldots\exp(L_{\chi_1}) J^{(n)} \nonumber \\
~~J_\phi&=&\exp(L_{\chi_{n}})\exp(L_{\chi_{n-1}})\ldots\exp(L_{\chi_1}) J^{(n)}_\phi \\
~u&=&\exp(L_{\chi_{n}})\exp(L_{\chi_{n-1}})\ldots\exp(L_{\chi_1}) u^{(n)}  \nonumber \\
~\phi&=&\exp(L_{\chi_{n}})\exp(L_{\chi_{n-1}})\ldots\exp(L_{\chi_1}) \phi^{(n)}\ ,\nonumber
\end{eqnarray}
where $L_{\chi}$ is the Poisson bracket operator, and $\exp(L_\chi)=\sum_{k=0}^\infty (1/k!)L_\chi^k$.
In practice, we truncate the latter sum at a maximum order $n_{max}>n$. We call $n_{max}$ the truncation order, and $n$ the normalization order.
Defining $H^{(0)}$ as the Hamiltonian \equ{eq:hambookkept}, the transformed Hamiltonian (\ref{hamnf}) is given by
$$
H^{(n)}=
\exp(L_{\chi_{n}})\exp(L_{\chi_{n-1}})...
\exp(L_{\chi_2})\exp(L_{\chi_1})H^{(0)}~~,
$$
and it contains the $n$-th order normal form as well as the first $n_{max}-n$ consecutive terms of the remainder series (\ref{rem}).
The method reduces now to determining the form of the functions $\chi_r$, $r=1,2,\ldots n$. This is accomplished by a recursive algorithm. Namely, after $r$ steps, we solve the {\em homological equation}:
$$
\left\lbrace Z_0, \chi_{r+1} \right\rbrace + \lambda^{r+1} h^{(r)}_{r+1} = 0\ ,
$$
where $Z_0 = J_{\phi} + \frac{1}{2} J$ is called the kernel of the normalization procedure, and $h^{(r)}_{r+1}$ is the set of all terms of $H^{(r)}_{r+1}$ whose Poisson bracket with $Z_0$ is different from zero. This yields $Z^{(r+1)}=H^{(r)}_{r+1}-h^{(r)}_{r+1}$. We refer the reader to the original articles of \cite{Hor1966} and \cite{Dep1969}, or \cite{CHR}, for a detailed presentation of the method of Lie series.

The choice of the functions $h^{(r)}$, $r=1,2,\ldots,n$ as above ensures that we avoid small divisors by making a resonant construction, in the sense that from the $h^{(r)}$ functions will
be excluded Fourier terms of the form $A(e,\delta,J) e^{i(k_1 u + k_2 \phi)}$ belonging to the resonant module:
$$
M = \left\lbrace \mathbf{k} \equiv (k_1,k_2) : k_1 \omega_2 + k_2 \omega_1 = 0 \right\rbrace\ ,
$$
where $\omega_1=1,\omega_2=1/2$ for the 2:1 secondary resonance. In the previous expression, the quantities $\omega_1, \omega_2$ denote the frequencies of the unperturbed part. Finally, the construction ensures that in every step we have advanced one order in the normal form construction, since the Hamiltonian
$$
H^{(r+1)}=\exp(L_{\chi_{r+1}}) H^{(r)}
$$
is normalized up to the order $r+1$, namely:
$$
H^{(r+1)}=Z_0+\lambda Z_{1}+\ldots+\lambda^{r} Z_{r} + \lambda^{r+1} Z_{r+1} + \lambda^{r+2} H^{(r+1)}_{r+2}+\ldots
$$
Unless explicitly required we hereafter skip superscripts from the notation of variables, and we simply indicate the order of normalization whenever needed.

In the rest of the paper, we present results by applying the above normalization scheme on the Hamiltonian (\ref{eq:hambookkept}) using a computer-algebraic program up to the normalization order $n=11$. The normalized Hamiltonian up to order 2 in the book-keeping (setting $\lambda=1$) reads
$$
H = Z_0 + Z_1 + Z_2 +\ldots
$$
with%
\beqano
Z_0 &=& \frac{1}{2} J + J_{\phi} \nonumber\\
Z_1 &=&  \delta J - \frac{3}{8}  e J \cos(2 u - \phi)\ \\
Z_2 &=&  \frac{89}{128} e^2 J- \frac{1}{4} J^2 - \frac{3}{4} e \delta J \cos(2 u - \phi)\ .
\eeqano
Next, we introduce another set of canonical variables for the resonant Hamiltonian:
$$
\phi\rightarrow\phi_F\ ,\qquad u \rightarrow \phi_{R}+\frac{1}{2} \phi_{F}\ ,\qquad J \rightarrow J_R\ ,\qquad J_{\phi} \rightarrow J_F - \frac{1}{2} J_R\ .
$$
The transformed Hamiltonian becomes:
$$
H = \delta J_R + \frac{89}{128} e^2 J_R - \frac{1}{4} J_R^2 - e J_R \left(\frac{3}{8}  + \frac{3}{4} \delta \right) \cos(2 \phi_R) +\ldots
$$
The action $J_F$ is now just a constant of the motion since its conjugate angle is not present in the Hamiltonian. The problem has been reduced to one degree of freedom and it is an integrable approximation of the original non-integrable system \eqref{HH21}.

We can further simplify the resonant Hamiltonian by applying a canonical transformation to Poincar\'e variables defined as
\begin{equation}\label{Pvar}
X = \sqrt{2 J_R} \sin{\phi_R}\ ,\quad Y = \sqrt{2 J_R} \cos{\phi_R}\ .
\end{equation}
Furthermore, since the integral $J_F$ still represents a dummy action variable, we can set $J_F=0$. One now gets the Hamiltonian in a polynomial form:
\begin{equation}\label{HXY11}
H = \frac{3}{16} e X^2 - \frac{89}{256} e^2 X^2 -
 \frac{1}{16} X^4 -  \frac{3}{16} e Y^2 - \frac{89}{256} e^2 Y^2 - \frac{1}{8} X^2 Y^2 -
 \frac{1}{16} Y^4 + \frac{1}{2}  X^2 \delta +
 \frac{3}{8} e X^2 \delta + \frac{1}{2} Y^2 \delta - \frac{3}{8} e Y^2 \delta +\ldots
\end{equation}

The complete form of the function of Eq. (\ref{HXY11}) up to order ${\cal O}(\lambda^5)$ is given in Appendix \ref{sec:AppendixNF}.

\subsection{Orbits and Phase Portraits}

\begin{figure}
    \includegraphics[width=\columnwidth]{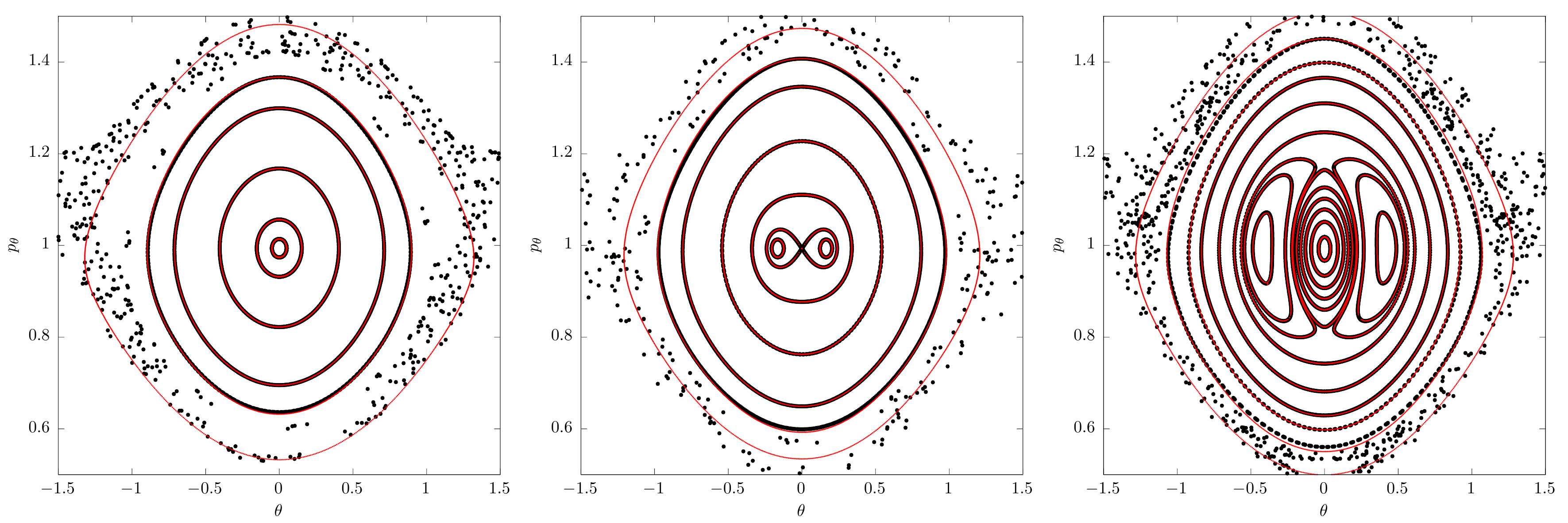}
    \caption{Poincar\'e surfaces of sections for different values of the control parameters $(\ep,e)$: left panel $(0.48,0.01)$, central panel $(0.5,0.01)$ and right panel $(0.52,0.01)$. The sections produced from the level curves of the resonant Hamiltonian normal form truncated at the normalization order 7 (red curves or gray in black and white version) are superposed to those produced from the numerical integration of the equations of motion (black points).}
    \label{fig:poincare_sections}
\end{figure}
The normal form of equation (\ref{HXY11}) provides an integrable approximation of our original system. By analyzing this function one can obtain valuable information about the original system. The solutions of the real system are encoded in the level curves of the integral $J_F=const$, or, equivalently, the constant energy curves of the Hamiltonian (\ref{HXY11}). In fact, by trivially integrating the orbits of \eqref{HXY11}, and back-transforming to the original variables via the transformation equations \eqref{lietra}, one obtains highly precise approximations of the solutions of the real system, at least in the domain of regular motions.
We now provide examples of this process, focusing on the bifurcation phenomena related to the 2:1 secondary resonance.

Figure~\ref{fig:poincare_sections} contains the relevant information. The black points correspond to numerically obtained phase space portraits for three different parameter values of $(\ep,e)$ selected so as to demonstrate the change of topology related to the stability of the central (synchronous) orbit. The original dynamical system described by equation \eqref{eq:soeom} has one degree of freedom (DOF) and an explicit time dependence. Its behavior is investigated by means of the computation of a stroboscopic Poincar\'e map. More precisely, we look at our solution every one orbit around the primary and record the values of $\theta$ and $p_\theta$ at each pericenter passage. The numerically obtained stroboscopic maps for a set of different initial conditions yield the portraits shown with black points in Fig.~\ref{fig:poincare_sections}. The chosen value of the eccentricity $e=0.01$ is low enough so that the phase space is dominated by regular motions, corresponding to closed invariant curves in the numerical phase portraits. However, some chaos can also be seen close to the pendulum separatrix-like limit of the synchronous resonant domain. The central fixed point is located at $\theta=0$ and $p_\theta\approx 1$ (varying with $(\ep,e)$). This is the point where the synchronous periodic orbit intersects the stroboscopic Poincar\'{e} section. Let us note that, in the original variables, the synchronous orbit is characterized by small periodic oscillations of the  angle $\theta$, whose amplitude increases, in general, with the values of both $(\ep,e)$. In physical terms, the spinning body does not show always exactly the same face to its attracting companion, since the orientation of the body relative to its orbital vector undergoes small periodic variations. An analytic formula for these variations is provided in Section~\ref{sec:po} below. On the other hand, as we move from left to right in Fig.~\ref{fig:poincare_sections} the chosen asphericity values are such as to depict the bifurcations undergone by the synchronous orbit. These happen close to the critical value
$\ep=0.5$, or $\delta=0$. We see that crossing this value by a little, turns the central synchronous orbit from stable to unstable. This bifurcation generates a new family of stable orbits (middle panel) along with their islands, delimited by a figure-eight shaped small internal separatrix. These are the islands of the 2:1 secondary resonance. However, for $\ep$ (or $\delta$) still larger (right panel), the central synchronous orbit has undergone one more transition from instability to stability, accompanied with a new bifurcation of an unstable pair of fixed points, changing once again the topology of the phase portrait.

Now, according to the normal form approximation, the invariant curves in the surfaces of section of Fig.~\ref{fig:poincare_sections}, should correspond also to the {\it level curves} of the one degree of freedom resonant Hamiltonian given by Eq. (\ref{HXY11}). We check this correspondence as follows: we compute first the constant energy level curves of the Hamiltonian (\ref{HXY11}) in the Poincar\'{e} variables $(X,Y)$ introduced in Eq. (\ref{Pvar}).
Then, using Eqs. \eqref{lietra}, we back-transform every point of one level curve into a point in the original variables $(\theta,p_\theta)$. To this end, we use the relation $J_F=0$, as well as the section condition $\phi_F=\phi=0$, implying that $X=\sqrt{2J^{(n)}}\sin u^{(n)}$, $Y=\sqrt{2J^{(n)}}\cos u^{(n)}$, where $n$ is the chosen normalization order. Finally, from the computed original variables $(J,u)$ we pass to $(\theta=u \mod 2\pi$, $p_\theta=1+J)$. The analytically found invariant curves (red curves), using the normalization order $n=7$, are superposed to the numerical phase portrait in Fig.~\ref{fig:poincare_sections}. Two remarks are relevant: i) we see that the analytical method is overall able to reproduce very accurately the numerical invariant curves for regular motions, and even to give a regular approximation to the shape of the separatrix-like thin chaotic layer surrounding the resonance. Most notably, ii) the analytical approximation precisely captures the change of the phase-space topology associated with the bifurcations of the central synchronous orbit, as well as the shape of the domain and the position of the stable 2:1 secondary resonance.

We now focus on a more detailed analytical study of these phenomena, and provide also error estimates and a specification of the limits of validity of the analytical method.

\subsection{Analytical approximation of the synchronous periodic orbit}\label{sec:po}
One of the most important solutions in the spin orbit problem is the periodic solution of the synchronous rotation itself. In this subsection, we take advantage of the simplified dynamics of the resonant normal form (\ref{HXY11}) in order to provide an explicit formula for the synchronous periodic orbit valid for quite high asphericity values ($\ep$ in the range $0.3 \sim 0.9$).

The synchronous periodic solution corresponds to the $X_0=0,Y_0=0$ equilibrium point of the polynomial normal form \eqref{HXY11}. Using the same procedure as described above for any other orbit, we back-transform this equilibrium point to a solution $\theta_{sync}(t;e,\delta)$ (and $p_{\theta,sync}(t;e,\delta) = \dot{\theta}_{sync}(t;e,\delta)$). Thus
\begin{equation}\label{eq:pothetaseries}
\theta_{sync}(t;e,\delta) = t + \Bigg(\exp{\left(L_{\chi_{n}}\right)}\ldots\exp{\left(L_{\chi_{2}}\right)}\exp{\left(L_{\chi_{1}}\right)} \left(\sqrt{\frac{2 J^{(n)}}{\omega_2}} \sin(u^{(n)})\right)\Bigg)\Bigg|_{J^{(n)}=0,u^{(n)}=t/2}
\end{equation}
\begin{equation}\label{eq:popthetaseries}
p_{\theta,sync}(t;e,\delta) = 1 + \Bigg(\exp{\left(L_{\chi_{n}}\right)}\ldots\exp{\left(L_{\chi_{2}}\right)}\exp{\left(L_{\chi_{1}}\right)} \left(\sqrt{2 J^{(n)}\omega_2} \cos(u^{(n)})\right)\Bigg)\Bigg|_{J^{(n)}=0,u^{(n)}=t/2}~~.
\end{equation}
The truncated series solution (\ref{eq:pothetaseries}) up to the book-keeping order 7 is given in Appendix \ref{sec:AppendixPO}, from which one easily obtains also the truncated series (\ref{eq:popthetaseries}) by direct time differentiation.

\begin{figure}
\includegraphics[width=\columnwidth]{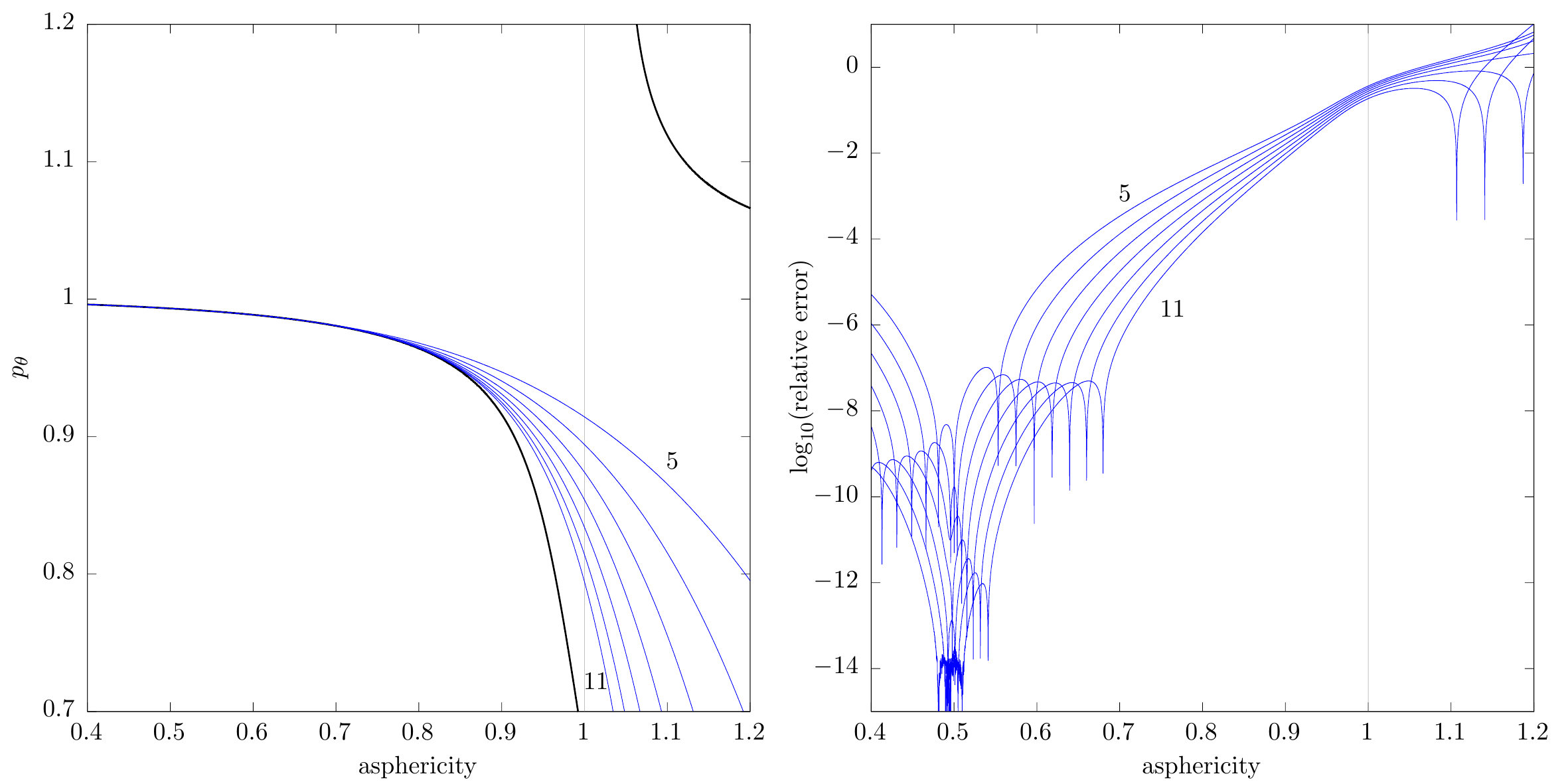}
    \caption{On the left panel: The characteristic curve of the periodic orbit corresponding to the 1:1 synchronous resonance for $e=0.01$. The characteristic curve of the $\beta$-mode is also shown on the top right of the plot. The numerically computed characteristic (black thick curve) is compared with those produced from the normal form truncated at different orders (from 5 top to 11 bottom (thin blue lines)). On the right panel: The relative error on the analytic computation of the position of the periodic solution. The grey line in both plots indicates the limiting value of $\ep$ beyond which the $\beta$-mode bifurcation takes place.}
    \label{fig:periodic_orbit}
\end{figure}

By further setting $t=0$ in equations (\ref{eq:pothetaseries}), (\ref{eq:popthetaseries}), one can obtain the position, on the Poincar\'e section, of the fixed point corresponding to the synchronous periodic orbit. One trivially finds $\theta(0)=0$, thus the only relevant parameter is the angular momentum $p_\theta(0)$ at $\theta=0$. Fixing one of the two parameters, say $e$, we can compute the {\it characteristic curve} yielding $p_\theta(0)$ as a function of the asphericity $\ep$, or the detuning $\delta$. The characteristic curve for $e=0.01$ is shown in the left panel of Fig.~\ref{fig:periodic_orbit}. The bold black curve starting at $\ep=0.4$ gives the characteristic curve by a numerical computation of the synchronous periodic orbit using the Newton-Raphson root finding method to locate the central fixed point in the stroboscopic Poincar\'{e} map with an accuracy threshold of $10^{-14}$. The analytical computation (via Eq. (\ref{eq:popthetaseries})) at different normalization orders, from $n=5$ to $n=11$, yields the thin curves superposed to the numerical characteristic curve in the same figure. The relative errors between the analytic and the numerical solutions are presented in the right panel of Fig.~\ref{fig:periodic_orbit}. We observe that the analytical computations at various normalization orders are all able to predict quite accurately (with errors of order $\sim 10^{-8}$ or smaller) the position of the synchronous fixed point up to an asphericity parameter $\ep\simeq 0.7$. From that point on, the error of the analytical computation increases. Near $\ep=0.9$, the 11th order approximation still yields a precision of about three significant digits. However, the approximations for all normalization orders $n$ collapse as we approach the value $\ep=1$. This collapse is associated with a new phenomenon which appears at this value of the asphericity, and causes the topology of the libration area around the synchronous state to change radically once more. This is a new tangential (or `saddle-node') bifurcation by which we have the birth of a pair of new synchronous solutions, one stable, known as the `$\beta$-mode' (see \cite{Melnikov}) and one unstable. A small part of the numerically computed characteristic of the $\beta$-mode solution is shown in the top right part of the left panel of Fig.~\ref{fig:periodic_orbit}. We note that the structure of the phase space near the bifurcation of the $\beta$-mode can be approximated by a different normal form approach (see, for example, \cite{Wis1984}). However, as it will be reported in a forthcoming work, it can also be recovered by our present method, using a different resonant module.

\subsection{Bifurcation Threshold of the 2:1 secondary resonance}\label{sec:bif11}

\begin{figure}
\includegraphics[width=0.5\columnwidth]{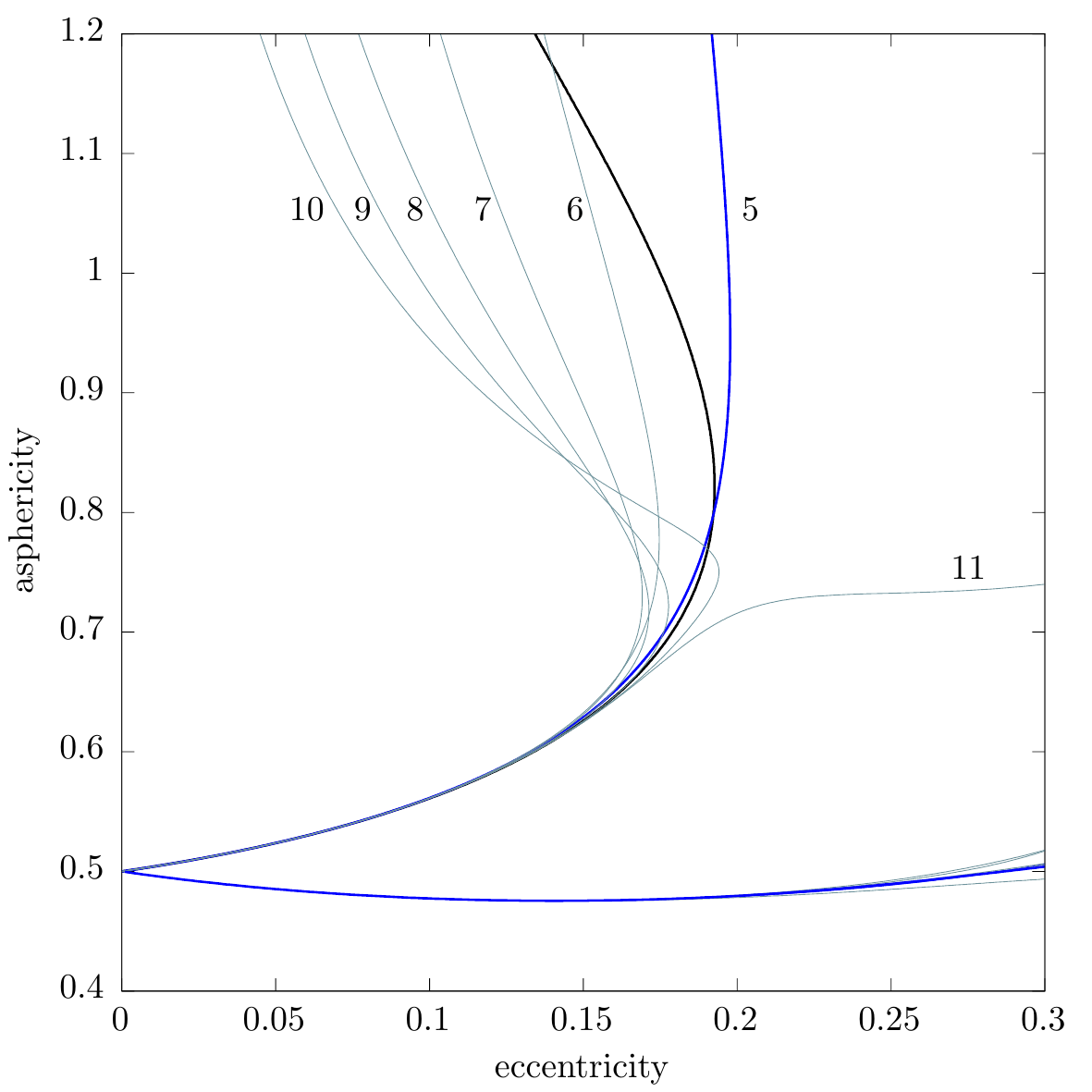}
    \caption{The bifurcation thresholds for the 2:1 secondary resonance. The numerically computed thresholds (lower and upper black thick lines) are compared with the thresholds computed from our analytic theory (blue lines) at various normalization orders indicated in the plot.}
    \label{fig:bifurcation}
\end{figure}
The determination of the bifurcation threshold of the 2:1 secondary resonance can be done numerically with a technique described in \cite{MS}. We compute the characteristic curve of the synchronous periodic orbit along $\ep$, or along $e$ (keeping, respectively, $e$ or $\ep$ fixed). We then compute the trace of the monodromy matrix $\mathcal{M}$. As long as $|Tr(\mathcal{M})|<2$ the family is stable. The bifurcation takes place at
$|Tr(\mathcal{M})|=2$. This method requires computing two nearby parameter values such that the bifurcation takes place in between. Bisection is then used to compute the bifurcation limit with greater accuracy.

Implementing the above technique we find that the bifurcation takes place at the loci shown as black curves in Fig.~\ref{fig:bifurcation}, which is essentially the same as in \cite{MS} (their figure 3). Note that, if we start moving upwards (along $\ep$), for $e$ fixed and small ($e<0.15$), we first encounter the lower bifurcation curve, where the central synchronous orbit turns from stable to unstable, while for larger $\ep$ we encounter the second bifurcation which restores the stability of the central orbit. This is in agreement with the transitions seen in Fig.~\ref{fig:poincare_sections} from left to right.

We now use our normal form technique to analytically determine these bifurcation thresholds as follows: from the normal form  in Poincar\'{e} variables Eq. \eqref{HXY11}, we compute the Hessian matrix at the origin
$$
K = \left( \begin{array}{cc}  \frac{\partial^2 H}{\partial X \partial Y} & \frac{\partial^2 H}{\partial Y^2} \\
    - \frac{\partial^2 H}{\partial X^2} & - \frac{\partial^2 H}{\partial Y \partial X}
  \end{array} \right)_{(X=0,Y=0)}\ .
$$
The eigenvalues $\lambda_1,\lambda_2$ of $K$ satisfy $\lambda_1+\lambda_2=0$. Then, the central synchronous orbit is stable if the eigenvalues are imaginary, and unstable if they are real. The bifurcation occurs whenever $\lambda_1=\lambda_2=0$. The Hamiltonian (\ref{HXY11}) contains only even powers of $X$ and $Y$, thus, the diagonal elements of $K$ are both equal to zero. Hence, a bifurcation occurs whenever one of the off-diagonal elements becomes also equal to zero. The two bifurcation curves are then computed as follows: i) the upper curve is computed from the condition
\begin{equation}\label{eq:upperlimit}
\left(\frac{\partial^2 H}{\partial Y^2}\right)_{(X=0,Y=0)}(\delta,e) = 0\ ,
\end{equation}
which is a polynomial algebraic equation in $(\delta,e)$. Similarly, the lower curve is computed from
\begin{equation}\label{eq:lowerlimit}
\left(\frac{\partial^2 H}{\partial X^2}\right)_{(X=0,Y=0)}(\delta,e) = 0\ .
\end{equation}
Explicit formulas of these algebraic equations are easily obtained from the form of $H$ (in the fifth order normal form approximation $Z^{(5)}$) as given in Appendix \ref{sec:AppendixNF}. In the last step, the algebraic equations (\ref{eq:upperlimit}) and (\ref{eq:lowerlimit}) are solved numerically for either $\ep$ or $e$ and a given value of the other parameter. We also tried to provide explicit formulas found by series inversion, but they prove to be less accurate than the ones found by the numerical solutions of the above algebraic equations.

In Fig.~\ref{fig:bifurcation} the numerically determined bifurcation threshold is compared with the one computed from our analytical theory. In the analytical computations we use the normal form formulas for all normalization orders from $n=5$ to $n=11$. It turns that the best approximation occurs for a value of $n$ changing along each of the bifurcation curves. This is a consequence of the asymptotic character of the normal form series, as detailed in the next subsection. Overall good results are found by the normal form at the normalization order $n=5$. As shown in Fig.~\ref{fig:bifurcation}, the $n=5$ analytical estimate is able to capture the turn around behavior of the upper bifurcation curve taking place at $e\approx 0.2$, $\ep\approx 0.75$. On the other hand, the whole lower curve is best represented analytically at the normalization order $n=11$. However, in this case too, the results at order $n=5$ are precise by several digits. Thus, the expressions up to order 5 given in Appendix A are sufficient for all practical purposes. We now turn our attention to a more detailed analysis of the method's errors .

\subsection{Error Analysis}
It has been mentioned already that, at any given normalization order $n$, the overall precision of the normal form method is related to the size of the remainder function $R^{(n)}$, which measures the distance between the normal form dynamics and the true dynamics. We now provide some explicit estimates of the precision of our method based on measuring the size of the remainder.

\begin{figure}
    \includegraphics[width=0.33\columnwidth]{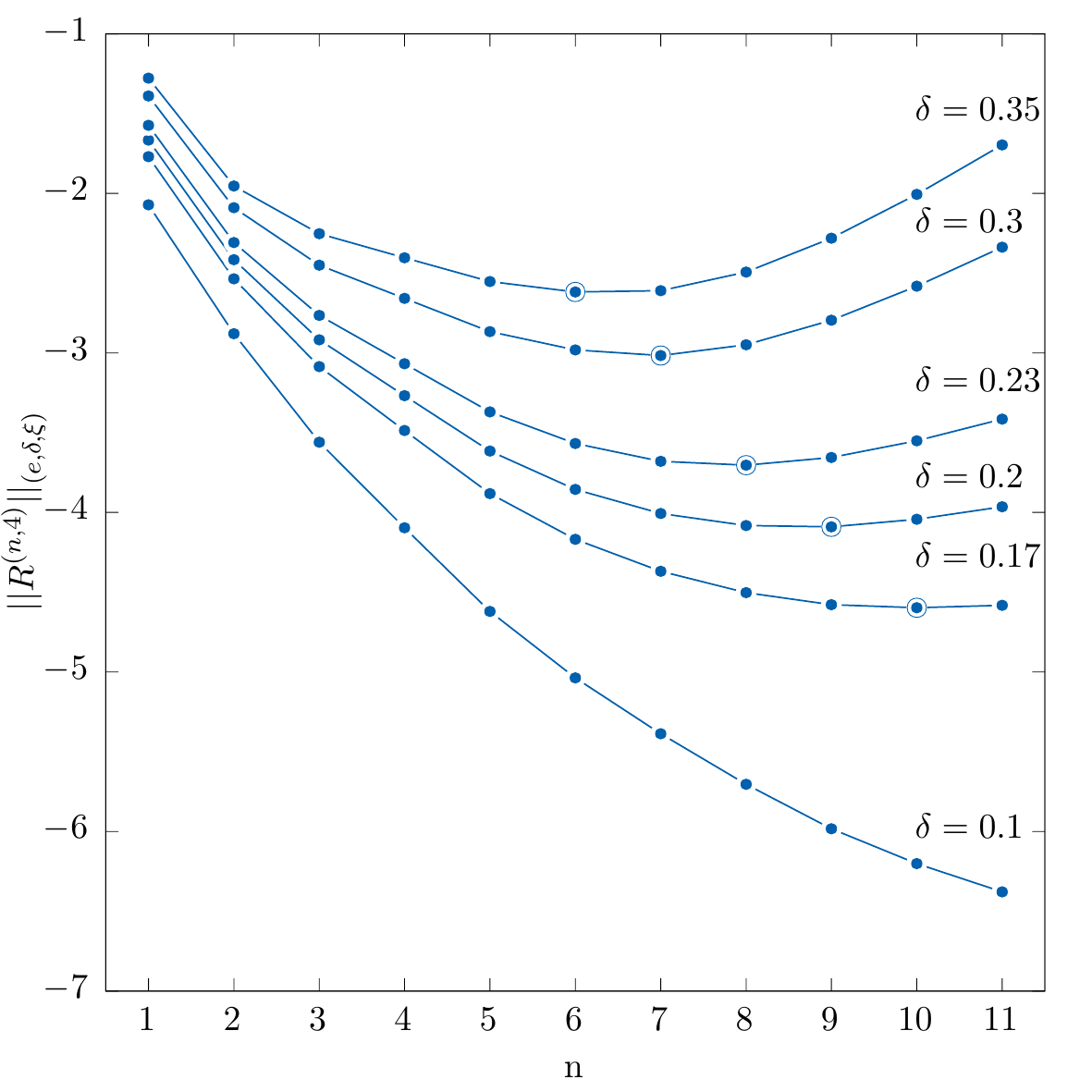}
    \includegraphics[width=0.33\columnwidth]{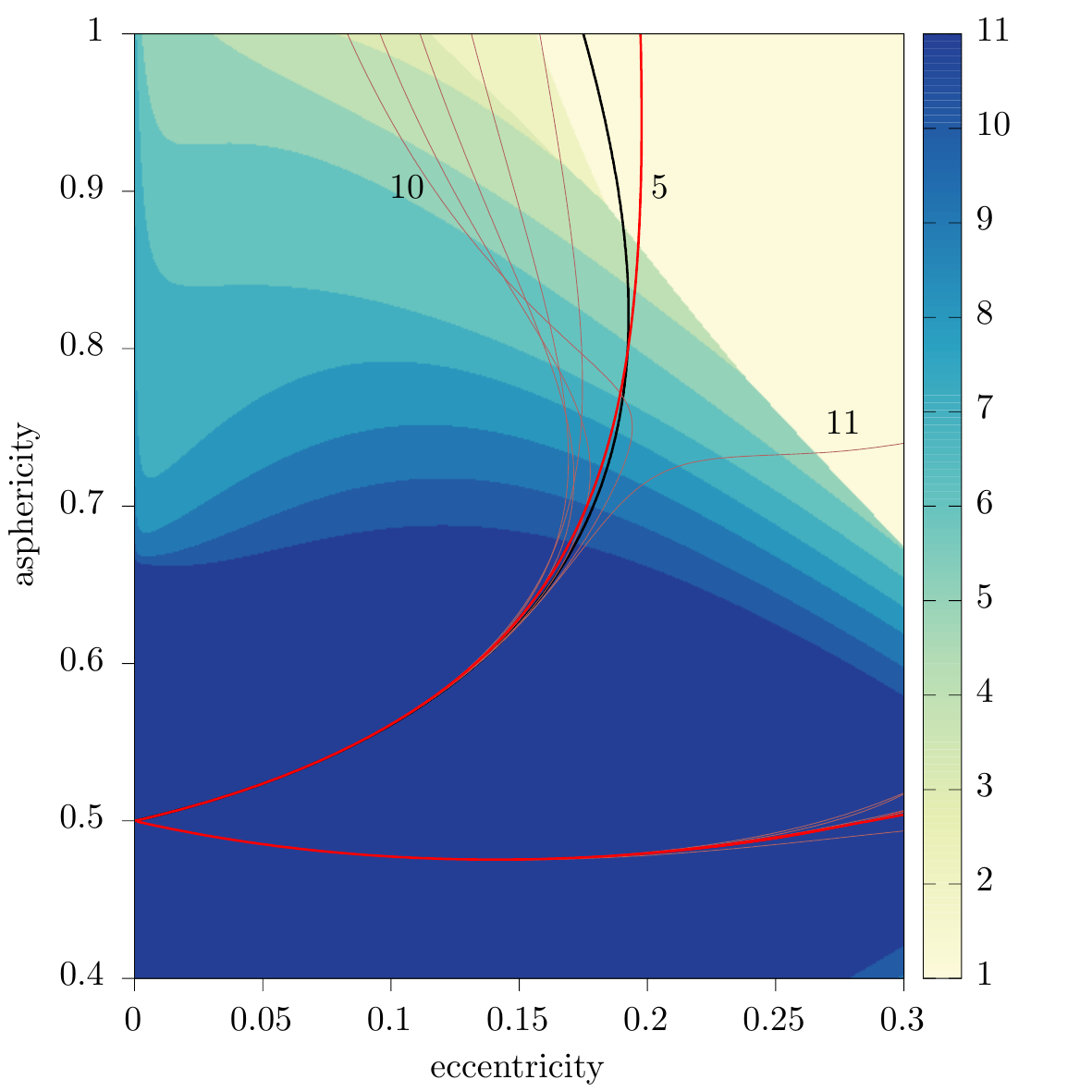}
    \includegraphics[width=0.33\columnwidth]{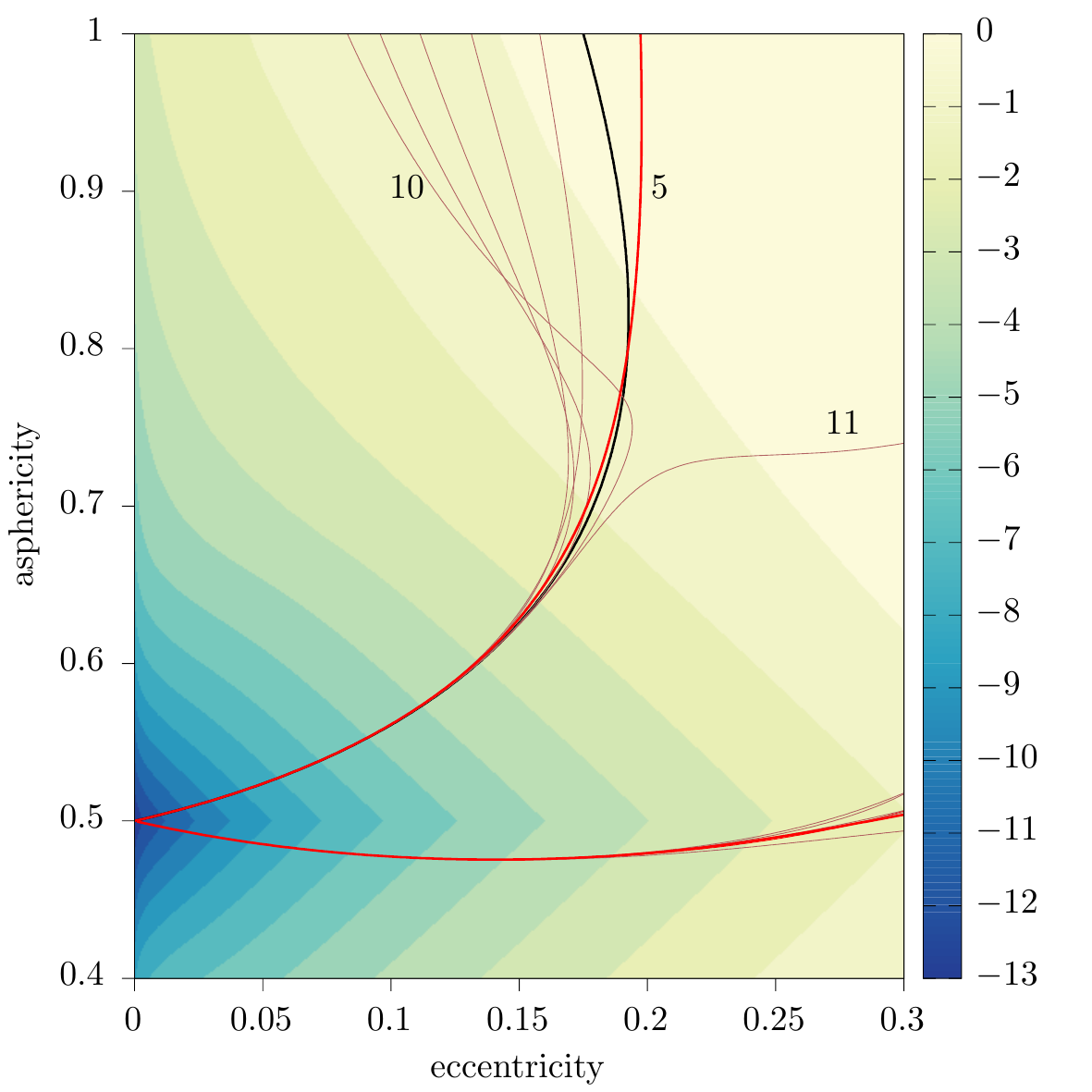}
    \caption{In the left panel: the norm of the remainder $|| R^{(n,4)} ||_{(e,\delta,\xi)}$ is plotted versus the normalization order $n$ for different values of $\delta$, reported in the plot, and $e$=0.03. The abscissa at the minimum of each curve (thick dots) determines the optimal order. In the central panel: the optimal order of the normal form $n_\textrm{opt}$ is computed in the parameter space $(\ep,e)$. In the right panel, the level of the numerical error (remainder value) is computed and shown in $log_{10}$ scale for the optimal order. In both color plots the numerical and analytical estimates for the  bifurcation thresholds are superposed.}
    \label{fig:error}
\end{figure}
The remainder has the form
\beq\label{remform}
R^{(n)}= \lambda^{n+1} H^{(n)}_{n+1} + \lambda^{n+2} H^{(n)}_{n+2}+\ldots\ = \sum_{s=1}^\infty \lambda^{n+s} \sum_{s_1,s_2,s_3,k_1,k_2} a^{(n,s)}_{s_1,s_2,s_3,k_1,k_2} e^{s_1} \delta^{s_2} J^{\frac{s_3}{2}} e^{i(k_1 u+k_2 \phi)}\ ,
\eeq
where $s_1,s_2,s_3 \in \positive$ and $k_1,k_2 \in \integer$, with lower and upper bounds depending on the order of the book-keeping $\lambda$. Since we cannot store infinitely many terms in the computer, we necessarily have to work with a truncated form of Eq.~(\ref{remform}). We define the $N$-term truncated remainder $R^{(n,N)}$ as
$$
R^{(n,N)} = \sum_{s=1}^N \sum_{s_1,s_2,s_3,k_1,k_2} a^{(n,s)}_{s_1,s_2,s_3,k_1,k_2} e^{s_1} \delta^{s_2} J^{\frac{s_3}{2}} e^{i(k_1 u+k_2 \phi)}\ .
$$
We also define a majorant norm, depending on $(e,\delta)$ as
\begin{equation}\label{eq:remnormtrunc}
|| R^{(n,N)} ||_{(e,\delta,\xi)} = \sum_{s=1}^{N} \sum_{s_1,s_2,s_3,k_1,k_2} \mid a^{(n,s)}_{s_1,s_2,s_3,k_1,k_2}\mid |e|^{s_1} |\delta|^{s_2} \xi^{\frac{s_3}{2}}~~.
\end{equation}
The positive quantity $\xi$ defines a disk $|J|<\xi$ around the origin, i.e., around the position of the central periodic orbit. Hence, equation (\ref{eq:remnormtrunc}) allows to estimate the size of the remainder in a domain around the origin. Basic theory (see, e.g., \cite{gior}) establishes that: i) the sequence $|| R^{(n,N)} ||_{(e,\delta,\xi)}$, for fixed $n,e,\delta,\xi$ and $N=1,2,\ldots$ is convergent for $e,\delta,\xi$ small enough. ii) The sequence $|| R^{(n,\infty)} ||_{(e,\delta,\xi)}$ is {\it asymptotic}. Namely, for $n$ small enough, the
quantity $|| R^{(n,\infty)} ||_{(e,\delta,\xi)}$ decreases as the normalization order $n$ increases, yielding the impression that the normal form procedure is convergent. However, beyond a certain order, called the {\it optimal order} $n_{opt}$, $|| R^{(n,\infty)} ||_{(e,\delta,\xi)}$ increases with $n$. Thus, the minimum possible value of the remainder, which corresponds also to the best normal form approximation, occurs at the normalization order $n=n_{opt}$. iii) Estimates based on Nekhoroshev theory (see e.g. \cite{CHR}) imply that $n_{opt}$ decreases as the small parameters (e.g. $e,\delta$ or $\xi$) increase.

We checked that all these properties are satisfied by the truncated remainders of our normal form construction. In fact, probing property (i) above, we found that the 4-term truncated remainder norm is sufficient to obtain nearly converged final values of the remainder in the whole investigated domain of the parameters ($e,\ep$).
On the other hand, the asymptotic behavior of the normal form series manifests itself in the size of the remainder, as shown in the left panel of Fig.~\ref{fig:error}, giving the size of the truncated remainder norm $||R^{(n,4)}||_{(e,\delta,\xi)}$ as a function of $n$ for different parameters $e,\delta$ as indicated in the caption. In these plots we set $\xi=0.01$. This value was selected by inspection of the phase portraits of Fig.~\ref{fig:poincare_sections}, which shows that the secondary resonance bifurcation phenomena take place in a disc of size $\sim 0.1$ around the origin, which translates to the action $J$ being of order $J\sim 0.1^2 = 0.01$. As shown in the left panel of Fig.~\ref{fig:error}, by increasing the parameter $\delta$ along a line of constant $e=0.03$, the optimal normalization order shifts to lower values, while the optimal remainder value becomes larger, i.e., the overall error of the method increases. On the other hand, close to $\delta=0$, the normal form computation is not optimal even at order 11. Repeating the computation of the optimal order and remainder in the whole domain of interest in the plane $(\ep=0.5+\delta,e)$, we arrive at the middle and right panels of Fig.~\ref{fig:error}, showing in color scale (or gray scale in printed format) the optimal order and the error estimates (optimal remainder value) as function of $(\ep,e)$. We note that the colour bands in the middle panel have a nearly horizontal structure, implying that the optimal order of our method is much more sensitive in the variation of the body's asphericity $\ep$ than in the orbital eccentricity $e$. On the other hand, from the right panel we conclude that the error goes beyond the second digit when $\ep$ surpasses the level of values $0.8 \sim 0.9$. These values provide the uppermost limit of applicability of the present method.

\section{A second example: The 2:1 secondary resonance in the 3:2 primary resonance}\label{sec:secondary32}
In order to further validate our method, in the present section we provide a second example of application, referring to the bifurcation of the secondary 2:1 resonance within the 3:2 primary one. Our analysis is nearly a repetition of previous steps, but with different parameter values. Starting again from the original spin-orbit Hamiltonian \eqref{eq:fullhamfour}, the transformation \eqref{rotransf1} reads as:
$$
p_1=p_\psi + \frac{3}{2}\ , \quad p_2=p_{\phi} - \frac{3}{2} p_{\psi}\ , \quad \psi = q_1 - \frac{3}{2} q_2\ , \quad \phi = q_2\ .
$$
The transformed Hamiltonian in the rotating frame reads as
$$
H=p_{\phi}+\frac{p^2_{\psi}}{2}-\frac{\ep^2}{4} \frac{7}{2}e \cos(2 \psi) - \frac{1}{4} \ep^2 \cos{(\phi+2 \psi)}+\frac{1}{8} e \ep^2 \cos{(2 \phi +\psi })+ H_{\text{pert}}(\psi,\phi;e,\ep)\ .
$$

\begin{figure}
    \includegraphics[width=0.5\columnwidth]{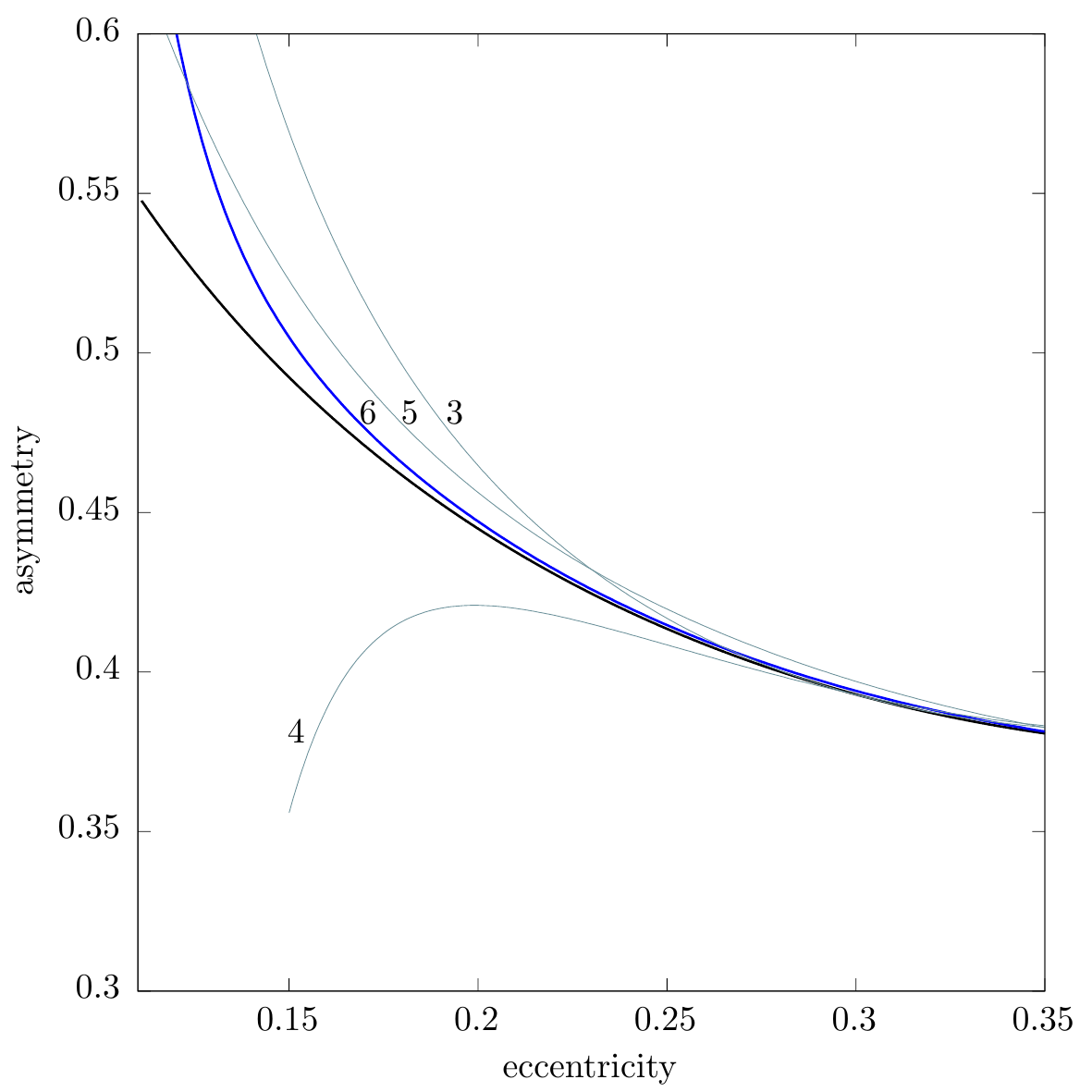}
    \caption{The bifurcation threshold of the 2:1 secondary resonance in the 3:2 primary. The numerically computed limit (black curve) is compared with those analytically determined for the resonant normal forms at orders $n=3,4,5,6$.}
    \label{fig:bif32}
\end{figure}
Notice here a key difference with respect to the 1:1 case: we observe that there are now low order terms, emanating from the harmonic coefficients $W(1,e)=1$ and $W(\frac{1}{2},e)=-\frac{e}{2}$, which are of the same,  or even {\it lower} order in the small parameters than the leading resonant term $\frac{\ep^2}{4} \frac{7}{2}e \cos(2 \psi)$. Since we are interested again in applying the detuning technique, it is useful to slightly change the book-keeping rules of Eqs. (\ref{booke}) - (\ref{bookJ}), in order to generate, at lowest order, a term with the appropriate harmonic frequency. We thus isolate the leading resonant term, and consider it of book-keeping order zero, while retain all remaining low order terms to order 1 in the book-keeping parameter $\lambda$. We emphasize that this choice is completely formal, and does not violate any of the rules of the normal form construction, which thereby proceeds in the usual way, i.e., in ascending powers of $\lambda$.

In more detail, Taylor-expanding, first, with respect to the resonant angle $\psi$ we obtain
$$
H=p_{\phi}+\frac{p^2_{\psi}}{2}+\frac{\ep^2}{2} \frac{7}{2}e \psi^2 + H_{\text{pert}}(\psi,\phi;e,\ep)\ .
$$
It is important to notice here that the frequency of the small amplitude oscillations is equal to $\ep \sqrt{\frac{7}{2}e}$, thus, it has a dependence on the eccentricity. Nevertheless, we can follow the procedure used before and define a detuning parameter $\delta$ as:
\begin{equation}\label{eq:32det}
\delta \equiv \ep \sqrt{\frac{7}{2}e} - \frac12\ .
\end{equation}
Note that the maximum possible asphericity value is $\ep_{max}=\sqrt{3}$, implying a lower bound in $e$ for the appearance of the secondary resonance $e_{min}=1/42$. Introducing \eqref{eq:32det}, the Hamiltonian takes the form:
$$
H=p_{\phi}+\frac{p^2_{\psi}}{2}+\frac{1}{8} \psi^2 +  H_{\text{pert}}(\psi,\phi;e,\delta)\ .
$$
Again, the integrable part of the Hamiltonian corresponds to a pair of a rotator and a harmonic oscillator with frequencies $\omega_1=1, \omega_2=1/2$ respectively. Introducing a set of action-angle variables for the integrable part according to \eqref{AARV}, we obtain
$$
H=J_{\phi}+ \frac{1}{2} J + H_{\text{pert}}(J,u,\phi;,e,\delta)\ .
$$
We finally introduce the book-keeping parameter $\lambda$. The resulting book-kept Hamiltonian becomes:
$$
H=J_{\phi}+ \frac{1}{2} J + \lambda H_1(J,u,\phi;e,\delta) + \lambda^2 H_2(J,u,\phi;e,\delta) + \ldots
$$
The Lie series normalization scheme used in Section~\ref{sec:norm} will also be used here to normalize the Hamiltonian. The kernel of the homological equation is $Z_0 = J_{\phi} + \frac{1}{2} J$. The normalization scheme is performed with resonant module
$$
M = \left\lbrace \mathbf{k} \equiv (k_1,k_2) : k_1 \omega_2 + k_2 \omega_1 = 0 \right\rbrace\ ,
$$
where $\omega_1=1,\omega_2=1/2$. The normalized Hamiltonian up to first order in book-keeping (setting $\lambda=1$) is given by
$$
H = Z_0 + Z_1
$$
with
\beqano
Z_0 &=& \frac{1}{2} J + J_{\phi}\nonumber\\
Z_1 &=&  J \delta - \left(\frac{J}{28 e}  + \frac{3}{14} e J \right) \cos{(2 u-\phi)}\ .
\eeqano
Introducing a set of resonant variables
$$
\phi \rightarrow 2 u - 2 \phi_{R}\ ,\quad J \rightarrow J_R\ ,\quad J_{\phi} \rightarrow J_F - \frac{1}{2} J_R\ ,
$$
the Hamiltonian becomes:
\begin{equation}\label{H3221}
H =  J_F + J_R \delta -\left(\frac{J_R}{28 e} +  \frac{3}{14} e J_R \right)  \cos{2 \phi_{R}} + \ldots
\end{equation}
The action $J_F$ is now, again, just a constant, since $\phi$ is an ignorable variable, and could be omitted.
For the analytic determination of the bifurcation limits it is convenient to introduce again the Poincar\'e set of variables
$$
X = \sqrt{2 J_R} \sin{\phi_R}\ ,\quad Y = \sqrt{2 J_R} \cos{\phi_R}\ ,
$$
which brings the Hamiltonian \eqref{H3221} in polynomial form as
$$
H =  \frac{X^2}{56 e} + \frac{3 e X^2}{28} - \frac{ Y^2}{56 e} -  \frac{3 e Y^2}{28} + \frac{X^2 + Y^2}{2}  \delta\ .
$$
Then, for the analytic determination of the bifurcation limits we follow the same procedure as in Section~\ref{sec:bif11}. The results are presented in Fig.~\ref{fig:bif32}, where the analytic limits produced from the  normalization orders $n=3,4,5,6$ are superposed to the numerically calculated limit. We observe that a good agreement between theory and numerical results is achieved after six normalization steps.

\section{Discussion and applications}\label{sec:discappl}

The analytical study of secondary spin-orbit resonances developed in this paper explores two normal form techniques, namely the `detuning' and the `book-keeping'. By a convenient combination of these techniques, we arrive at very accurate analytical results which represent well the bifurcation sequences of secondary resonances in a parameter space containing the orbital eccentricity $e$ and the body's asphericity $\ep$. Here we provided examples of application in the case of the 2:1 secondary resonance, focusing on its bifurcation from the most important primary resonance, i.e., the synchronous one, and giving also a short second example referring to the 3:2 primary resonance. Our main conclusions are the following:

1) Relatively low order (5 or 7) truncations of the resonant normal form series constructed as above are able to reproduce the structure of the phase portraits around the synchronous resonance in a domain of regular librations extending up to the separatrix-like chaotic layer surrounding the primary resonance. In particular, they allow to follow analytically the islands formed by the bifurcation of secondary resonances as well as the latters' position in phase space.

2) We used a normal form stability analysis to compute analytically the bifurcation thresholds of the secondary resonances, recovering these thresholds with several significant figures.

3) We give analytic formulas representing the synchronous primary periodic orbit, as well as the bifurcation thresholds of the 2:1 secondary resonance.

4) We provided an analysis of the errors of the method based on the asymptotic behavior of the remainder of the normal form series. We computed the optimal normalization orders and showed their agreement with expectations from the theory of normal forms. We give precise maps indicating the degree of accuracy of the analytical solutions in the parameter space $(\ep,e)$.

\begin{figure}
    \centering
    \includegraphics[width=0.5\columnwidth]{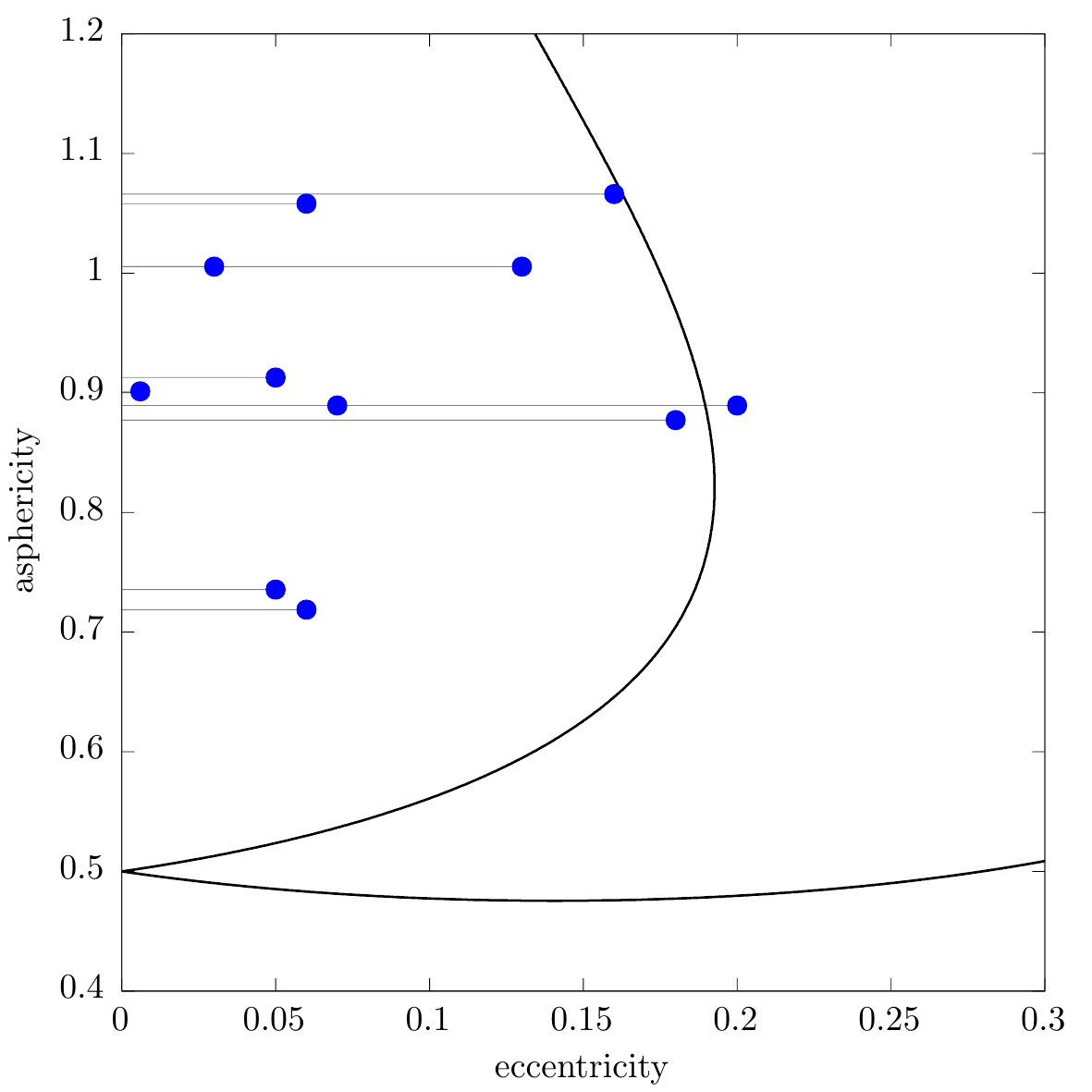}
    \caption{The  positions of the 11 synchronous binary asteroids systems are superposed on the 2:1 secondary resonance threshold curve. The data are taken from Pravec (2016). The eccentricity values only correspond to 3-$\sigma$ upper bounds.}
    \label{fig:binarydata}
\end{figure}
We emphasize that the method is algorithmically quite convenient, and easily transcribable to a computer-algebraic program. In fact, it requires some standard computations, i.e., the computation of a high
order expansion of the orbital solution in the eccentricity, the introduction of a detuning parameter to measure the distance from the exact resonance and a suitable book-keeping of terms in the normal form construction.
The proposed technique gives results in full agreement with the numerical predictions and it can be extended
to any primary resonance as well as to any secondary resonance around a primary resonance.

The importance of analytical theories for secondary resonances in the spin-orbit dynamics stems from the fact that the latters' study has led to physically relevant conclusions, like the tidal heating due to a forced secondary resonance libration as it was proposed for Enceladus in \cite{Wisdom}. The occurrence of secondary resonances strongly depends on the values of the parameters, in particular the eccentricity and the asphericity. The corresponding bifurcation diagram gives information on the stability of the primary resonance and therefore poses some constraints on the satellites which cannot rotate, e.g., synchronously, since the 1:1 primary resonance is unstable (\cite{MS}). As an example, Melnikov and Shevchenko, with the help of an empirical law that estimates $\ep$ with respect to the radius of small body, have placed all the minor planetary satellites on the $(\ep,e)$-plane suggesting that many of them will never reach the synchronous rotation since, for their physical parameters, there exists no stable synchronous solution. More recently, \cite{Pravec} published a list of physical parameters for binary asteroids observed in the synchronous state. In Fig.~\ref{fig:binarydata}, we superpose these data to the bifurcation plot for the 2:1 secondary resonance. Note here that the depicted values of the eccentricities only correspond to 3-$\sigma$ upper bounds. Yet, all but one of the observed objects appear in the region delimited by the upper bifurcation curve of the 2:1 secondary resonance, i.e. they  probably all lie within the domain in which the synchronous resonance is stable.

Examples as the above indicate that an analytical theory of secondary resonances may provide useful constraints on the domains where stable spin-orbit resonant configurations of minor bodies would be expected to be observed.

\section*{Acknowledgments}
We thank G. Voyatzis for useful suggestions. A.C. was partially supported by PRIN-MIUR 2010JJ4KPA$\_$009, GNFM-INdAM and by the European MC-ITN grant Stardust. I.G. was supported by the European MC-ITN grant Stardust. G.P. was partially supported by the European MC-ITN grant Stardust and GNFM-INdAM.


\clearpage
\appendix

\section{5th order Normal Form in Poincar\'e variables}\label{sec:AppendixNF}
The $n$-th order normal form of Eq. (\ref{HXY11}), expressed in Poincar\'e variables, has the form
$$
H=Z^{(n)}=\sum_{i_1,i_2,i_3,i_4,i_5} a_{i_1,i_2,i_3,i_4,i_5}
\lambda^{i_1} e^{i_2} \delta^{i_3} X^{i_4} Y^{i_5}
$$
with exponents $i_j$, $j=1,\ldots,5$, determined by the book-keeping rules of Eqs. \equ{booke}, \equ{bookd}, \equ{bookJ}. The entire form of $Z^{(5)}$ is given in the table \ref{tab:z5}1.
\begin{table}
    \centering
    \label{tab:z5}
    \begin{minipage}[b]{.40\columnwidth}

    \begin{tabular}{cccccc} 
        \hline
        $i_1$ & $i_2$ & $i_3$ & $i_4$ & $i_5$ & $a_{i_1,i_2,i_3,i_4,i_5}$ \\
        \hline
1   &   1   &   0   &   2   &   0   &   $3/16$  \\
1   &   1   &   0   &   0   &   2   &   $-3/16$ \\
1   &   0   &   1   &   2   &   0   &   $1/2$   \\
1   &   0   &   1   &   0   &   2   &   $1/2$   \\
2   &   2   &   0   &   2   &   0   &   $-89/256$   \\
2   &   0   &   0   &   4   &   0   &   $-1/16$ \\
2   &   2   &   0   &   0   &   2   &   $-89/256$   \\
2   &   0   &   0   &   2   &   2   &   $-1/8$  \\
2   &   0   &   0   &   0   &   4   &   $-1/16$ \\
2   &   1   &   1   &   2   &   0   &   $3/8$   \\
2   &   1   &   1   &   0   &   2   &   $-3/8$  \\
3   &   2   &   0   &   2   &   0   &   $-1/3$  \\
3   &   3   &   0   &   2   &   0   &   $-365/4096$ \\
3   &   1   &   0   &   4   &   0   &   $-3/64$ \\
3   &   2   &   0   &   0   &   2   &   $-1/3$  \\
3   &   3   &   0   &   0   &   2   &   $365/4096$  \\
3   &   1   &   0   &   0   &   4   &   $3/64$  \\
3   &   2   &   1   &   2   &   0   &   $-223/256$  \\
3   &   2   &   1   &   0   &   2   &   $-223/256$  \\
3   &   1   &   2   &   2   &   0   &   $-3/8$  \\
3   &   1   &   2   &   0   &   2   &   $3/8$   \\
4   &   2   &   0   &   2   &   0   &   $-1/18$ \\
4   &   3   &   0   &   2   &   0   &   $-31/240$   \\
4   &   4   &   0   &   2   &   0   &   $62221/196608$  \\
4   &   2   &   0   &   4   &   0   &   $3619/10240$    \\
4   &   0   &   0   &   6   &   0   &   $-1/64$ \\
4   &   2   &   0   &   0   &   2   &   $-1/18$ \\
4   &   3   &   0   &   0   &   2   &   $31/240$    \\
4   &   4   &   0   &   0   &   2   &   $62221/196608$  \\
4   &   2   &   0   &   2   &   2   &   $-2981/5120$    \\
4   &   0   &   0   &   4   &   2   &   $-3/64$ \\
4   &   2   &   0   &   0   &   4   &   $3619/10240$    \\
4   &   0   &   0   &   2   &   4   &   $-3/64$ \\
4   &   0   &   0   &   0   &   6   &   $-1/64$ \\
4   &   2   &   1   &   2   &   0   &   $-22/9$ \\
4   &   3   &   1   &   2   &   0   &   $239/256$   \\
4   &   1   &   1   &   4   &   0   &   $15/64$ \\
4   &   2   &   1   &   0   &   2   &   $-22/9$ \\
4   &   3   &   1   &   0   &   2   &   $-239/256$  \\
4   &   1   &   1   &   0   &   4   &   $-15/64$    \\
4   &   2   &   2   &   2   &   0   &   $-3/256$    \\

        \hline
    \end{tabular}

    \end{minipage}\quad
    \begin{minipage}{.40\columnwidth}

    \begin{tabular}{cccccc} 
        \hline
        $i_1$ & $i_2$ & $i_3$ & $i_4$ & $i_5$ & $a_{i_1,i_2,i_3,i_4,i_5}$ \\
        \hline
4   &   2   &   2   &   0   &   2   &   $-3/256$    \\
4   &   1   &   3   &   2   &   0   &   $13/16$ \\
4   &   1   &   3   &   0   &   2   &   $-13/16$    \\
5   &   3   &   0   &   2   &   0   &   $-3/160$    \\
5   &   4   &   0   &   2   &   0   &   $186527/115200$ \\
5   &   5   &   0   &   2   &   0   &   $-458595/1048576$   \\
5   &   2   &   0   &   4   &   0   &   $23/120$    \\
5   &   3   &   0   &   4   &   0   &   $-15443/409600$ \\
5   &   1   &   0   &   6   &   0   &   $-57/1024$  \\
5   &   3   &   0   &   0   &   2   &   $3/160$ \\
5   &   4   &   0   &   0   &   2   &   $186527/115200$ \\
5   &   5   &   0   &   0   &   2   &   $458595/1048576$    \\
5   &   2   &   0   &   2   &   2   &   $-37/60$    \\
5   &   1   &   0   &   4   &   2   &   $-57/1024$  \\
5   &   2   &   0   &   0   &   4   &   $23/120$    \\
5   &   3   &   0   &   0   &   4   &   $15443/409600$  \\
5   &   1   &   0   &   2   &   4   &   $57/1024$   \\
5   &   1   &   0   &   0   &   6   &   $57/1024$   \\
5   &   2   &   1   &   2   &   0   &   $-19/27$    \\
5   &   3   &   1   &   2   &   0   &   $2213/14400$    \\
5   &   4   &   1   &   2   &   0   &   $148928/85027$  \\
5   &   2   &   1   &   4   &   0   &   $36991/19200$   \\
5   &   0   &   1   &   6   &   0   &   $1/32$  \\
5   &   2   &   1   &   0   &   2   &   $-19/27$    \\
5   &   3   &   1   &   0   &   2   &   $-2213/14400$   \\
5   &   4   &   1   &   0   &   2   &   $148928/85027$  \\
5   &   2   &   1   &   2   &   2   &   $73457/19200$   \\
5   &   0   &   1   &   4   &   2   &   $3/32$  \\
5   &   2   &   1   &   0   &   4   &   $36991/19200$   \\
5   &   0   &   1   &   2   &   4   &   $3/32$  \\
5   &   0   &   1   &   0   &   6   &   $1/32$  \\
5   &   2   &   2   &   2   &   0   &   $-208/27$   \\
5   &   3   &   2   &   2   &   0   &   $6989/4096$ \\
5   &   1   &   2   &   4   &   0   &   $-49/128$   \\
5   &   2   &   2   &   0   &   2   &   $-208/27$   \\
5   &   3   &   2   &   0   &   2   &   $-6989/4096$    \\
5   &   1   &   2   &   0   &   4   &   $49/128$    \\
5   &   2   &   3   &   2   &   0   &   $39/64$ \\
5   &   2   &   3   &   0   &   2   &   $39/64$ \\
5   &   1   &   4   &   2   &   0   &   $-15/32$    \\
5   &   1   &   4   &   0   &   2   &   $15/32$ \\
\hline
    \end{tabular}
    \end{minipage}
\caption{}
\end{table}

\section{7th order Synchronous Periodic Solution }\label{sec:AppendixPO}
Setting $\lambda=1$, the analytical formula for the synchronous periodic orbit takes the form
$$
\theta_{sync}(t)=t+ \sum_{i_1,i_2,k} a_{i_1,i_2,k} e^{i_1} \delta^{i_2} \sin(k t).
$$
Then, $p_{\theta,sync}$ can be computed as $d\theta_{sync}(t)/dt$. The coefficients of the solution $\theta_{sync}(t)$, up to seventh order, are given in table \ref{tab:sync7}.
\begin{table}
    \centering
    \caption{}
    \label{tab:sync7}
    \begin{minipage}{.40\columnwidth}
    \begin{tabular}{cccc} 
        \hline
        $i_1$ & $i_2$ & $k$ & $a_{i_1,i_2,k}$ \\
        \hline
1   &   0   &   1   &   -0.6666666666666666 \\
3   &   0   &   1   &   3.1104938271604925  \\
5   &   0   &   1   &   -14.032182980599643 \\
7   &   0   &   1   &   19.705429188285102  \\
1   &   1   &   1   &   -3.5555555555555554 \\
3   &   1   &   1   &   29.195720164609046  \\
5   &   1   &   1   &   -186.12448294280674 \\
1   &   2   &   1   &   -8.296296296296296  \\
3   &   2   &   1   &   152.25147873799722  \\
5   &   2   &   1   &   -539.1273886736492  \\
1   &   3   &   1   &   -15.802469135802468 \\
3   &   3   &   1   &   545.9295868312755   \\
1   &   4   &   1   &   -32.13168724279835  \\
3   &   4   &   1   &   732.4495644444443   \\
1   &   5   &   1   &   -63.912208504801086 \\
1   &   6   &   1   &   -128.0585276634659  \\
2   &   0   &   2   &   -0.35   \\
4   &   0   &   2   &   1.741652557319223   \\
6   &   0   &   2   &   -6.925800246883582  \\
2   &   1   &   2   &   -1.8488888888888884 \\
4   &   1   &   2   &   15.763010262870566  \\
6   &   1   &   2   &   -48.23782824357248  \\
2   &   2   &   2   &   -4.238222222222222  \\
4   &   2   &   2   &   74.21459558818941   \\
2   &   3   &   2   &   -7.944217283950616  \\
    \hline
    \end{tabular}
    \end{minipage}
\begin{minipage}{.40\columnwidth}
\begin{tabular}{cccc} 
        \hline
        $i_1$ & $i_2$ & $k$ & $a_{i_1,i_2,k}$ \\
        \hline
4   &   3   &   2   &   140.3971755301135   \\
2   &   4   &   2   &   -16.101325432098765 \\
2   &   5   &   2   &   -31.977028126200267 \\
3   &   0   &   3   &   -0.3612522045855381 \\
5   &   0   &   3   &   1.9630691996904597  \\
7   &   0   &   3   &   -4.36611960712759   \\
3   &   1   &   3   &   -2.155303602922651  \\
5   &   1   &   3   &   18.173301226683666  \\
3   &   2   &   3   &   -6.448019311569428  \\
5   &   2   &   3   &   50.85174061588128   \\
3   &   3   &   3   &   -15.546704845546174 \\
3   &   4   &   3   &   -20.06780604724124  \\
4   &   0   &   4   &   -0.41432949105568173    \\
6   &   0   &   4   &   2.3057242238485145  \\
4   &   1   &   4   &   -2.6340673553371943 \\
6   &   1   &   4   &   14.649998958089611  \\
4   &   2   &   4   &   -8.266419946523065  \\
4   &   3   &   4   &   -12.425566492716285 \\
5   &   0   &   5   &   -0.5055804692808663 \\
7   &   0   &   5   &   2.268623489193706   \\
5   &   1   &   5   &   -3.298490290414931  \\
5   &   2   &   5   &   -7.585427848029809  \\
6   &   0   &   6   &   -0.6294608630807215 \\
6   &   1   &   6   &   -3.480595010536174  \\
7   &   0   &   7   &   -0.7236805357989855 \\
    \hline
    \end{tabular}
    \end{minipage}
\end{table}


\bsp    
\label{lastpage}
\end{document}